\documentclass[preprint,preprintnumbers,amsmath,amssymb]{revtex4}
\usepackage{amsfonts}    
\usepackage{amssymb}
\usepackage{latexsym}
\usepackage{graphicx}
\usepackage{rotating}
\usepackage{multirow}
\usepackage[usenames,dvipsnames]{color}
\usepackage[active]{srcltx}

\usepackage{hyperref}
\usepackage{color}

\newcommand{\al}{\alpha}
\newcommand{\pa}{\partial}

\newcommand{\veps}{\varepsilon}

\newcommand{\la}{\lambda}

\newcommand{\rar}{\rightarrow}

\begin{document}

\title{Anharmonic oscillator: a solution}
\date{\today}

\author{Alexander~V.~Turbiner}
\email{turbiner@nucleares.unam.mx}
\author{J.C.~del~Valle}
\email{delvalle@correo.nucleares.unam.mx}
\affiliation{Instituto de Ciencias Nucleares, Universidad Nacional
Aut\'onoma de M\'exico, Apartado Postal 70-543, 04510 M\'exico,
D.F., Mexico{}}

\begin{abstract}
It is shown that for the one-dimensional quantum anharmonic oscillator with potential
$V(x)= x^2+g^2 x^4$ the Perturbation Theory (PT) in powers of $g^2$ (weak coupling regime)
and the semiclassical expansion in powers of $\hbar$ for energies coincide.
It is related to the fact that the dynamics in $x$-space and in $(gx)$-space corresponds to the same energy spectrum with effective coupling constant $\hbar g^2$. Two equations, which govern the dynamics in those two spaces, the Riccati-Bloch (RB) and the Generalized Bloch (GB) equations, respectively, are derived. The PT in $g^2$ for the logarithmic derivative of wave function leads to PT (with polynomial in $x$ coefficients) for the RB equation and to the true semiclassical expansion in powers of $\hbar$ for the GB equation, which corresponds to a loop expansion for the density matrix in the path integral formalism. A 2-parametric interpolation of these two expansions leads to a uniform approximation of the wavefunction in $x$-space with unprecedented accuracy $\sim 10^{-6}$ locally and unprecedented accuracy $\sim 10^{-9}-10^{-10}$ in energy for any $g^2 \geq 0$. A generalization to the radial quartic oscillator is briefly discussed.
\end{abstract}

\maketitle
\noindent
{\bf Introduction.} It is common knowledge that the quartic anharmonic oscillator
\begin{equation}
\label{AHOQ}
      V(x)\ =\ x^2+ g^2 x^4\ =\ \frac{1}{g^2}\,\hat{V}(gx)\ ,
\end{equation}
where $\hat V(u) = u^2 + u^4$, which was one of the first problems tackled by newly the born quantum mechanics, appears practically in all branches of physics. It reveals a great richness of non-trivial properties reflecting a complexity of Nature. Needless to say that enormous efforts were dedicated to the exploration of this problem. One particular direction of study was the field-theoretical treatment of (\ref{AHOQ}) as $(0+1)$ quantum field theory.

In a seminal paper by Bender and Wu \cite{BW:1969} the theory of the weak coupling regime was created, where the coupling constant $g^2$ is small. It was shown that in this domain the problem is, in fact, algebraic (practically everything can be calculated by linear algebra means) (i), that there exist infinitely-many square-root branch points in the complex $g^2$ plane of energy accumulating to $g^2=0$ (the so-called {\it horn structure} of singularities; recently, their existence was proved rigorously in \cite{EG:2009}) as a reflection of the Landau-Zener theory of level crossings (ii), and that the perturbation theory
\begin{equation}
\label{EPT}
   E\ =\ \sum_0^{\infty}\ a_n\, g^{2n}\ ,
\end{equation}
has zero radius of convergence (iii). Many year after \cite{TU:1988} a sort of duality between weak and strong coupling regimes was discovered,
\begin{equation}
\label{ESC}
    E\ =\ g^{\frac{2}{3}}\sum_0^{\infty} b_n g^{-\frac{4n}{3}}\ ,
\end{equation}
where the coefficient $b_n$ can be represented as an infinite sum of the $a_k, k=0,1,2,\ldots$ coefficients and visa versa. Based on this property the first nine coefficients $b_n, n=0,1,\ldots 8$ were calculated.

The problem of bound states in non-relativistic quantum mechanics is governed by the time-independent Schr\"odinger equation
\begin{equation}
\label{SE}
  {\cal H}\ =\ -\frac{\hbar^2}{2 m} \pa_{x}^2 + V(x)\quad ,\quad {\cal H}\Psi=E\Psi\quad ,\quad \int |\Psi|^2 d x < \infty\ ,\quad \pa_x \equiv \frac{d}{dx}\ ,
\end{equation}
where $V$ is potential (\ref{AHOQ}). Take an exponential representation for the wavefunction
\begin{equation}
\label{exp}
  \Psi\ =\ e^{-\frac{1}{\hbar}\phi} \ ,
\end{equation}
and substitute it into (\ref{SE}). We arrive at the well-known Riccati equation
\begin{equation}
\label{Riccati}
  \hbar\,y' \ -\  y^2\ =\ E\ -\ x^2 - g^2 x^4 \ , \quad y\,=\,\phi'\ ,
\end{equation}
where for the sake of convenience $m=1/2$. It contains $\hbar$ in front of the leading derivative term and $g^2$ in front of the anharmonic term: it leads to a bifurcation in $\hbar$ and to the divergence of the perturbation theory in powers of $g^2$. Now let us get rid off the explicit $\hbar$ dependence in this equation. There are two ways to do so.

\noindent
$\bf (I)$\ {\it Riccati-Bloch equation}. Let us introduce in (\ref{Riccati}) a new variable $v$, function $\mathcal{Y}$ and energy,
\begin{equation}
\label{RB-variables}
  x\ =\ {\hbar^{1/2}}\,{v} \ ,\ y\ =\ \hbar^{1/2}\, \mathcal{Y}\left(v\right)\ ,\ E\ =\ {\hbar}\,\veps\ ,
\end{equation}
respectively, and also the effective coupling
\begin{equation}
\label{coupling}
   \la \ =\ \hbar^{1/2}\,g\ .
\end{equation}
We arrive at the so-called {\it Riccati-Bloch} (RB) equation
\begin{equation}
\label{RB}
  \pa_v {\cal Y}\ -\ {\cal Y}^2\ =\ \veps (\la^2)\ -\ v^2 - \la^2\, v^4\ , \quad \pa_v \equiv \frac{d}{dv}\ ,\ v\in (-\infty,\infty)\ .
\end{equation}
This equation has no $\hbar$-dependence: {\it it describes dynamics in a ``quantum", $\hbar$-dependent coordinate $v$ (\ref{RB-variables}) instead of $x$, which is governed by the $\hbar$-dependent, effective coupling constant $\la$ (\ref{coupling}).} If we develop the Perturbation Theory (PT) for the ground state, see e.g. \cite{T:1984},
\begin{equation}
\label{PT-veps}
   \veps\ =\ \sum_0^{\infty} \la^{2n} \veps_n\ ,\quad \ \veps_0\ =\ 1\ ,\ \veps_1\ =\ \frac{3}{4} \ ,\ \ldots
\end{equation}
\begin{equation}
\label{PT-Y}
   {\cal Y}\ =\ \sum_0^{\infty} \la^{2n} {\cal Y}_n(v)\ ,\quad \ {\cal Y}_0\ =\ v\ ,\ {\cal Y}_1\ =\ \frac{v}{2}(v^2 + \frac{3}{2})\ ,\ \ldots \ ,
\end{equation}
where ${\cal Y}_n=vP_n(v^2)$ with $P_n$ as an $n$th degree polynomial,
it becomes clear that (\ref{PT-veps}) is simultaneously the perturbation series in powers of $g^2$ and the semiclassical expansion in powers of $\hbar$, since the coefficients $\veps_n$ are numbers. Contrary to that, the expansion (\ref{PT-Y}) is the PT expansion in powers of $g^2$ only, since the corrections ${\cal Y}_n(v)$ are $\hbar$-dependent. Hence, (\ref{PT-Y}) is not semiclassical expansion in powers of $\hbar$. Expansion (\ref{PT-Y}) mimics the asymptotic expansion at small $v$ in Eq.(\ref{RB}),
\begin{equation}
\label{Y-asymp}
   {\cal Y}\ =\ \veps\, v + \frac{\veps^2-1}{3}\, v^3 \ +\ \ldots \ ,
\end{equation}
where $\veps$ is given by (\ref{PT-veps}).

\noindent
$\bf (II)$ {\it Generalized Bloch equation}. Let us introduce in (\ref{Riccati}) a new variable $u$, function $\mathcal{Z}$ and energy,
\begin{equation}
\label{GB-variables}
u\ =\ {g\,x}\ =\ \la v\ ,\quad y\ =\ \frac{1}{g}\mathcal{Z}(u)\ ,\quad E\ =\ {\hbar}\,\veps\ ,
\end{equation}
keeping the same effective coupling constant (\ref{coupling})
\[
   \la \ =\ \hbar^{1/2}\,g\ ,
\]
in the Riccati equation (\ref{Riccati}) assuming $g \neq 0$. We arrive at
\begin{equation}
\label{GB}
   \la^2\,\pa_u\mathcal{Z}(u)\ -\ \mathcal{Z}^2(u)\ =\ \la^2\,\veps(\la^2)\ -\ u^2-u^4 \quad ,\quad \pa_u\equiv\frac{d}{du}\ ,\quad u \in (-\infty,\infty)\ ,
\end{equation}
cf.(\ref{RB}). This is the so-called {\it the Generalized Bloch equation} (GB), see e.g. \cite{Shuryak:2018}, it requires a regularization at $\la \rar 0$, which will lead to the RB equation. This equation describes dynamics in {\it classical ($\hbar$-independent) coordinate} $u=g\,x$. Now we develop a PT in powers of $\la$ in (\ref{GB}). It is evident that the expansion of the energy $\veps$ in powers of $\la$ (\ref{PT-veps}) remains the same as in the RB equation (\ref{RB}), 
\begin{equation*}
   \veps\ =\ \sum_0^{\infty} \la^{2n} \veps_n\ ,
\end{equation*}
unlike the expansion for ${\mathcal{Z}}$,
\begin{equation}
\label{PT-Z}
   {\cal Z}\ =\ \sum \la^{2n} {\cal Z}_n(u)\ ,\ {\cal Z}_0\ =\ u{\sqrt{1 + u^2}}\ ,\ {\cal Z}_1\ =\ \frac{{\cal Z}^{\prime}_0-\veps_0}{2{\cal Z}_0}\ ,\quad \ldots \ ,
\end{equation}
cf.(\ref{PT-Y}), $\veps_0=1$. Purely algebraically, one can calculate ${\cal Z}_{2,3,4,5,6,\ldots}$ - any finite number of corrections. It can be immediately recognized that ${\cal Z}_0$ is, in fact, the classical momentum at zero energy and $\la=1$, ${\cal Z}_1$ is related with the derivative of the logarithm of the determinant. In general, the expansion (\ref{PT-Z}) is the true semiclassical expansion in powers of  $\hbar$, as well as the expansion in powers of $g^2$. Integrating $\int {\cal Z}\,du$ (and putting $\hbar=1$) we arrive at the expansion of the phase,
\begin{equation}
\label{PT-Z-phi}
   \phi\ =\ \frac{1}{3g^2}\,(1+g^2 x^2)^{3/2} + \frac{1}{4} \log(1+g^2 x^2) + \left(2n + p + \frac{1}{2}\right) \log[1+(1 + g^2 x^2)^{1/2}] \ +\ \ldots \ ,
\end{equation}
where $n=p=0$ for the ground state \footnote{The first term was calculated in \cite{T:1984} under the name {\it leading log approximation in quantum mechanics}.}. Remarkably, higher order corrections ${\cal Z}_{2+i}, i=0,1,2,\ldots$ in (\ref{PT-Z}) do not generate logarithmic terms. For $(2n+p)$-excited state with quantum number $2n$ and parity $p=0,1$, the function ${\cal Z}$ contains $(2n+p)$ simple poles with residues $(-1)$ in addition to the non-singular part. In the case of excited states the coefficient in front of the second logarithmic term is the state-dependent. Expansion (\ref{PT-Z-phi}) mimics the asymptotic expansion at large $u=gx$ in Eq.(\ref{GB}),
\begin{equation}
\label{PT-Z-asymp}
   \phi\ =\ \frac{g}{3}\,x^2\,|x|\ +\ \frac{1}{2 g}\,|x|\ +\ \left(n + \frac{p}{2} + \frac{1}{2}\right) \log( x^2) \ +\ O\left(\frac{1}{|x|}\right) \ .
\end{equation}
The first three terms are universal: they grow as $|x| \rar \infty$ and do not depend on energy $\veps$, thus, on the state considered. Furthermore, the first two terms emerge from the expansion of the classical action, the first term in the expansion (\ref{PT-Z-phi}). 

\bigskip

\noindent
$\bf (III)$\ {\it Approximating eigenfunctions}.  Without loss of generality we put $\hbar=1$, hence, $v=x$ and the RB equation (\ref{RB}) coincides with the Riccati equation (\ref{Riccati}). Let us match the expansions (\ref{Y-asymp}) and (\ref{PT-Z-phi}) in the exponent of (\ref{exp}) and multiply it by a polynomial which carries the information about the nodes. As a result we arrive at the following function for the $(2n+p)$-excited state with quantum numbers $(n,p)$, $n=0,1,2,\ldots\ ,\ p=0,1$\,,
\[
 \Psi^{(n,p)}_{(approximation)}\ =\
 \frac{x^p P_{n,p}(x^2; g^2)}{\left(B^2\ +\ g^2\,x^2 \right)^{\frac{1}{4}}
 \left({\al B}\ +\ \sqrt{B^2\ +\ g^2\,x^2} \right)^{2n+p+\frac{1}{2}}}
\]
\begin{equation}
\label{final}
   \exp \left(-\ \dfrac{A\ +\ (B^2 + 3)\,x^2/6\ +\ g^2\,x^4/3}
  {\sqrt{B^2\ +\ g^2\,x^2}} \ +\ \frac{A}{B}\right)\ ,
\end{equation}
where $P_{n,p}$ is some polynomial of degree $n$ in $x^2$ with positive roots, $\al=1$. Here $A=A_{n,p}(g^2),\ B=B_{n,p}(g^2)$ are two parameters of interpolation. If $\al=0$, the expression (\ref{final}) becomes the one which was found in \cite{TURBINER:2005,TURBINER:2010} by interpolating the expansions (\ref{Y-asymp}) and (\ref{PT-Z-phi}).
In the limit $g^2 \rar 0$, the function (\ref{final}) becomes the one of the harmonic oscillator,
\[
   \Psi^{(0)}\ =\ x^p L_n^{(-\frac{1}{2}+\frac{p}{2})}(x^2) \ e^{-\frac{x^2}{2}}  \ ,
\]
where $L^{(-\frac{1}{2}+\frac{p}{2})}_n$ is the Laguerre polynomial with index $\al=-\frac{1}{2}+\frac{p}{2}$.

The simplest way to find the parameters $A,B$ is to consider (\ref{final}) as a variational trial function moving subsequently from the ground state to higher excited states. In order to realize this program: (i) we have to calculate the effective potential,
\begin{equation}
\label{e-potential}
  V^{(n,p)} - E_0 \ =\ \frac{\pa_x^2\, \Psi^{(n,p)}}{\Psi^{(n,p)}}\quad ,
\end{equation}
where $V^{(n,p)}(x=0)=0$, then (ii) impose the orthogonality conditions
\[
   \langle \Psi^{(k,p)}\,,\,\Psi^{(n,p)}  \rangle =0\ ,\ k=0,1,2\ldots (n-1) \ ,
\]
assuming that the functions $\Psi^{(k,p)}, k=0,1,2\ldots (n-1)$ are already found, and (iii) calculate
the variational energy,
\begin{equation}
\label{Evar}
  E^{(n,p)}_{var}\ =\ \min_{\{A,B\}} \, \left[E_0 + \langle \Psi^{(n,p)}\,(V - V^{(n,p)})\,\Psi^{(n,p)}  \rangle\right]\ ,
\end{equation}
which is the expectation value of the deviation of the original potential (\ref{AHOQ}) from the effective potential (\ref{e-potential}). In the framework of the non-linearization procedure \cite{T:1984} one can develop a fast-convergent perturbation theory with respect to $(V - V^{(n,p)})$.

It can be shown that in the small and large $g^2$ limits, the parameters $A,B$ behave like
\[
     A^{(n,p)} \rar \frac{1}{3g^2} \quad ,\quad A^{(n,p)} \rar - a^{(n,p)}\, g^{2/3}\ ,
\]
respectively, while
\[
     B^{(n,p)} \rar 1 \quad ,\quad B^{(n,p)} \rar  b^{(n,p)}\, g^{2/3}\ .
\]
Note that the functions $a^{(n,p)}, b^{(n,p)} > 0$ are very smooth, they can be easily fitted with relative accuracy $\sim 10^{-3}$,
\begin{equation*}
	 a_{fit}^{(n,p)}= (8.869\ +\ 23.120\,N\ + 7.856\,N^2\ +\ 4.140N^3\ +\ 0.262\,N^4 )^{1/3}
\ ,\ b_{fit}^{(n,p)} \ =\ (10.040\ +\ 3.255 N)^{1/3}\ ,
\end{equation*}
where $N=2n+p$, realizing the dependence on the quantum numbers.

\begin{figure}[h]
	\includegraphics[width=10cm]{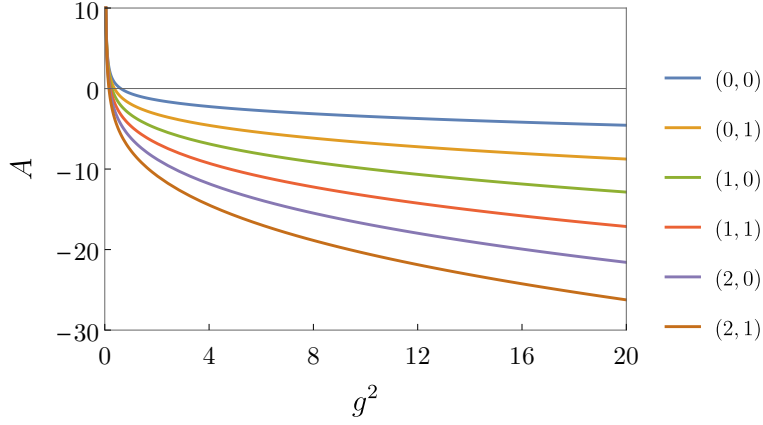}
	\caption{Optimal  parameter $A$  as function of the coupling constant $g^2$ for the first six low-lying states.}
	\label{Aplot}
\end{figure}

\begin{figure}[h]
	\includegraphics[width=10cm]{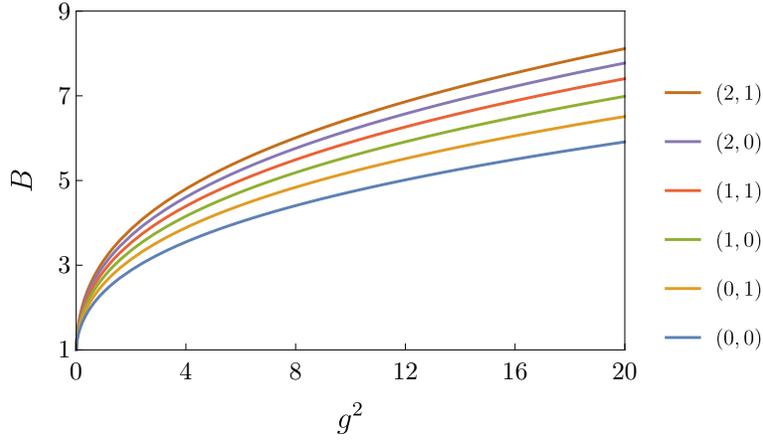}
	\caption{Optimal  parameter $B$  as function of the coupling constant $g^2$ for the first six low-lying states.}
	\label{Bplot}
\end{figure}

\begin{figure}[h]
	\centering
	\includegraphics[width=10cm]{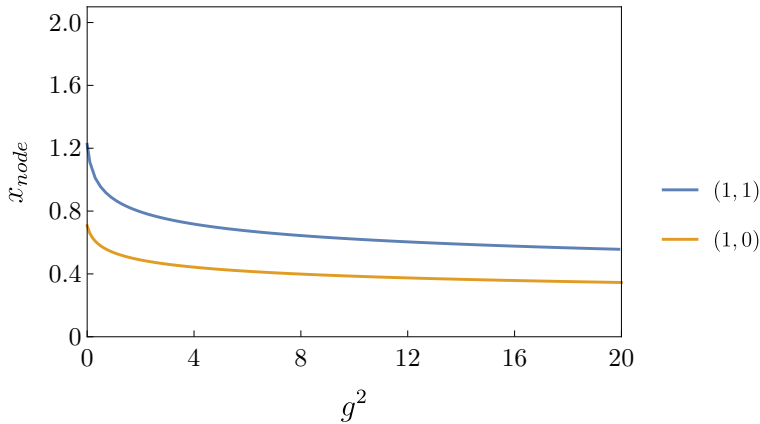}
	\caption{Positive nodes $x_{node} > 0$ \ {\it vs}\ $g^2$ for the states $(1, 0)$, $(1, 1)$.}
	\label{NodesPlot}
\end{figure}

Concrete numerical calculations were carried out for the six lowest states $(n,p)$: $n=0,1,2$ and $p=0,1$  for the $g^2=0.1,\ 1,\ 10,\ 20,\ 100$. As a result the variational energy was found with relative accuracy $\sim 10^{-10}$ - it was confirmed by making accurate calculations in the Lagrange mesh method \cite{Baye:2015} with 50 mesh points (giving relative accuracy $\sim 10^{-13}$), see also \cite{Tur-delValle:2021}. Plots of parameters $A,B$ are presented in Figs.\ref{Aplot}-\ref{Bplot}.
Surprisingly, in order to reach the above-mentioned accuracy it was enough to find $A,B$ with 4 significant digits only. In Fig.\ref{NodesPlot} the behavior of nodes $x_{node}$ for $(n,p), n=1$ and $p=0,1$ are shown. By introducing a new coupling constant
\[
     {\tilde\la}(g^2)\ =\ (0.008 + g^2)^{1/3}\quad ,\quad {\tilde\la}(0)\ = \ 0.2\ ,
\]
one can interpolate the parameters $A,B$ for all the above-mentioned six states in a form,
\[
    A\ =\ \frac{1}{g^2\,\tilde\la} P_5 (\tilde\la)\quad ,\quad B\ =\ \frac{Q_3(\tilde\la)}{Q_2(\tilde\la)}\ ,
\]
where $P_5, Q_3, Q_2$ are polynomials of 5,3,2 degrees, respectively, with different coefficients.
In particular, for the ground state
\[
\hskip -0.5cm	A^{(0,0)}\ =\ \frac{-0.0171\ +\ 0.4205\,\tilde\la\ -\ 0.1990\,\tilde\la^{2}\ +
      \ 1.039 \,\tilde\la^{3}\ -\ 0.0567\, \tilde\la^{4}\ -\ 1.797\,\tilde\la^{5}}{g^2\,\tilde\la} \ ,
\]
\[
\hskip -0.5cm	B^{(0,0)}\ =\  \frac{0.3716\ +\ 5.476\,\tilde\la\ +\ 2.231\,
      \tilde\la^2+33.51\,\tilde\la^3}{1\ +
      \ 0.9981\,\tilde\la\ +\ 15.61\,\tilde\la^2}\ .
\]
Use of these interpolated parameters in (\ref{final}) leads to relative accuracy in energy $\sim 10^{-9}$ (or less). Similar accuracy is reached for the all six studied states.

It is interesting to investigate the polynomials $P_{n,p}$ in (\ref{final}) which carry the information about non-zero nodes of the eigenfunction,
\[
   P_{n,p}\ =\ 1\ -\ a_2^{(n,p)}\,x^2\ +\ a_4^{(n,p)}\,x^4\ +\ \ldots\ +\ a_{2n}^{(n,p)}\,(-x^2)^n \ .
\]
For the harmonic oscillator with $V=x^2$ (it corresponds to $g^2 \rar 0$, the weak coupling regime) 
\begin{align*}
 a_2^{(n,p)}\ &=\  \frac{4\,n}{(p+1)(p+2)} \ ,\\
 a_4^{(n,p)}\ &=\  \frac{16\,n(n-1)}{(p+1)(p+2)(p+3)(p+4)}\ ,
\end{align*}
\[
  \ldots
\]
\[
    a_{2n}^{(n,p)}\ =\ \frac{\Gamma(\frac{1+p}{2})}{\Gamma(\frac{1+p}{2}+n)} \ ,
\]
which emerge from the properties of the Laguerre polynomials, while at the ultra-strong coupling regime,
$g^2 \rar \infty$, thus for the potential $V=x^4$, the first two coefficients can be fitted as
\begin{itemize}
	\item[$\bullet$] $p=0$
\begin{align*}
	a_{2,fit}^{(n,0)}\ &=\ n\left(7.372\ +\ 10.872\,n\right) ^{1/3}\ ,\\
	a_{4,fit}^{(n,0)}\ &=\ n\,(n-1)\left(0.830\ +\ 1.420\,n\right)^{2/3}\ ,
\end{align*}
with relative error $\lesssim 0.5\%$ for any $n \in [0, 40]$ and
	\item[$\bullet$] $p=1$
\begin{align*}
	a_{2,fit}^{(n,1)}\ &=\ n\left(0.834\ +\ 0.773\,n\right)^{1/3}\ ,\\
	a_{4,fit}^{(n,1)}\ &=\ n\,(n-1)\left(0.167\ +\ 0.127\,n\right)^{2/3}\ ,
\end{align*}
with relative error $\lesssim 0.12\%$ for any $n \in [0, 40]$.
\end{itemize}

In the non-linearization procedure using the multiplicative perturbation theory \cite{T:1984} one can calculate analytically the first corrections to (\ref{final}),
\[
    \Psi\ =\ \Psi^{(n,p)}_{(approximation)}(1 - \phi_1 - \phi_2 \ldots)\ .
\]
It can be shown numerically that for all six studied states at any $g^2 \in [0, \infty)$, the first correction
\begin{equation}
\label{dphi}
  |\phi_1| \lesssim 10^{-6} \ ,\ \forall x \in (-\infty , \infty)\ .
\end{equation}
For all six studied states the correction $\phi_1$ can be found explicitly in closed analytic form and we believe that (\ref{dphi}) can be verified analytically. Besides that in the non-linearization procedure the corrections $E_2, E_3$ to energy (\ref{Evar}) can be also calculated analytically and then verified numerically. They indicate an extremely high rate of convergence for the energy, $10^{-3}$ -  $10^{-4}$ in agreement with the Lagrange mesh calculations, see \cite{Baye:2015}, \cite{Tur-delValle:2021}, with 75 mesh points for the $g^2=0.1,\ 1,\ 10,\ 20,\ 100$, where the relative accuracy $10^{-20}$ in energy is reached easily
\footnote{Needless to say that the Lagrange mesh method in many occasions allows to reach easily the extremely high accuracy. Out of curiosity it was checked that with 1000 mesh points the energy of the states from $N=0$ to $N=80$ for the $g^2=0.1,\ 1,\ 10,\ 20,\ 100$ can be found with more than 140 figures with $\sim 250$ seconds of CPU time with single processor of 2.6 GHz for each value of $g^2$ and each state, see \cite{Tur-delValle:2021}.}.

\noindent
{\bf (IV) \it Radial quartic anharmonic oscillator.}\ Formalism developed in Sections {\bf (I)-(III)} can be extended to the case spherically-symmetric $D$-dimensional quartic anharmonic oscillator with potential
\[
     V\ =\ r^2\ +\ g^2\ r^4\ ,
\]
cf.(\ref{AHOQ}), where $r^2=\sum_{i=1}^D x_i^2$. Separating out in the Schr\"odinger equation the angular variables we arrive at the familiar radial Schr\"odinger equation, which can be converted to the Riccati equation. In turn, introducing the quantum and classical coordinates similar to those used for one-dimensional case the Riccati equation can be transformed to either Riccati-Bloch or the Generalized Bloch equations, which are similar to (\ref{RB}) and (\ref{GB}) equations, respectively. In latter two equations the same effective coupling constant $\la$ as in one-dimensional case (\ref{coupling}) occurs. It allows to construct the PT and semiclassical expansions in the same way as was made earlier. In fact, these expansions look similar to (\ref{PT-Y}) and (\ref{PT-Z}) after replacement $x \rar r$ (and replacement $1 \rar D$ in degree of one of the prefactors). Matching these two expansions we arrive at
\[
	\Psi^{(n_r,\ell)}_{(approximation)}\ =\
	\frac{r^\ell P_{n_r}(r^2)}{\left(B^2\ +\ g^2\,r^2 \right)^{\frac{1}{4}}
		\left({ B}\ +\ \sqrt{B^2\ +\ g^2\,r^2} \right)^{2n_r+\ell+\frac{D}{2}}}
\]
\begin{equation}
		\exp \left(-\ \dfrac{A\ +\ (B^2 + 3)\,r^2/6\ +\ g^2\,r^4/3}
		{\sqrt{B^2\ +\ g^2\,r^2}} \ +\ \frac{A}{B}\right)\ ,
\label{Approximant_radial}
\end{equation}
cf.(\ref{final}), where $\hbar=1$ and $P_{n_r,\ell}$ is some polynomial of degree $n_r$ in $r^2$ with positive roots; here $(n_r,\ell)$ are radial and angular quantum numbers, $n_r=0,1,2,\ldots$ and $\ell=0,1,2,\dots$. The parameters $A=A_{n_r,\ell}(g^2),\ B=B_{n_r,\ell}(g^2)$ are two parameters of interpolation, they can be found variationally, in particular, for the ground state by making minimization of the expectation value of the radial Hamiltonian. The results for the ground state for different $D$ and $g^2$ are presented in Table I. All digits for energies in Table I are exact: it was checked in Lagrange mesh calculations, also by calculating the first correction to the variational energy $E_2$ and making comparison with variational energies obtained with more parametric generalization of (\ref{Approximant_radial}), see \cite{delValle}, Part II. Details will be presented elsewhere as well as a generalization for the excited states.

\begin{table}[h]
\caption{Ground state energy for the quartic radial potential $V(r)=r^2\,+\,g^2\,r^4$ for
          $D=1,2,3,6$ and $g^2=0.1, 1, 10$, calculated using (\ref{Approximant_radial}) in variational calculations.}
	{\setlength{\tabcolsep}{0.3cm}		
\begin{tabular}{|c|cccc|}
\hline
	$g^2$&$D=1$&$D=2$&$D=3$&$D=6$ \\ \hline
		\rule{0pt}{4ex}
		0.1      &1.065\,285\,509\,54&2.168\,597\,211\,3   &3.306\,872\,013\,2  &6.908\,332\,111\,2      \\[4pt]
		1.0      &1.392\,351\,642 &2.952\,050\,092       &4.648\,812\,704    &10.390\,627\,296      \\[4pt]
		10.0    &2.449\,174\,07  &5.349\,352\,82 &8.599\,003\,46  &19.936\,900\,4     \\[4pt]
\hline
\end{tabular}}
\end{table}

\bigskip

\noindent
{\bf Conclusions.}\ As a results of a straightforward interpolation for the logarithm of the wavefunction of the Taylor expansion at small distances and the true semiclassical expansion valid for the large distances, approximate eigenfunctions for the quartic anharmonic oscillator (\ref{final}) are built for arbitrary coupling constant.
They lead to unprecedented accuracies for both the eigenvalues and eigenfunctions. Formula (18) manifests the approximate solution of the problem of the quartic anharmonic oscillator. 
This formula admits straightforward generalization (22) to the case of radial quartic anharmonic oscillator.
Since the generalized Bloch equation (\ref{GB}) and the true semiclassical expansion (\ref{PT-Z}) can be written for any anharmonic oscillator bounded from below, the approximate eigenfunctions of type (\ref{final}) can be built. Evidently, similar procedure can be realized for any radial anharmonic oscillator, see \cite{delValle}.

\vspace{-5mm}
\section*{Acknowledgments}
\vspace{-3mm}

The authors thank C.M.~Bender, L.~Brink, V.~Korepin, A.A.~Migdal, N.A.~Nekrasov, A.M.~Polyakov, M.A.~Shifman, B.~Simon, R.~Schrock, A.I.~Vainshtein, M.~Znojil and, 
especially, E.~Shuryak for their interest to the work, useful comments and discussions.
This work is partially supported by CONACyT grant A1-S-17364 and DGAPA grant IN113819 (Mexico).



\begin{thebibliography}{99}

\bibitem{BW:1969}
         C.M.~Bender, T.T.~Wu,\\
         {\it Anharmonic Oscillator},\\
         {\it Phys Rev  \bf 184},  1231-1260 (1969)

\bibitem{EG:2009}
         A.~Eremenko A.~Gabrielov,\\
         {\it Analytic continuation of eigenvalues of a quartic oscillator},\\
         {\it Comm. Math. Phys. \bf 287}, 431-457 (2009)

\bibitem{TU:1988}
         A.V.~Turbiner, and  A.G.~Ushveridze,\\
         {\it Anharmonic oscillator: Constructing the strong coupling expansions},\\
         {\t J. Math. Phys. \bf 29}, {2053-2063} (1988)

\bibitem{T:1984}
         A.V.~Turbiner,\\
          {\it Soviet Phys. - Usp. Fiz. Nauk. \bf 144}, 35-78 (1984),\\
          {\it Sov. Phys. Uspekhi \bf 27}, 668-694 (1984) (English Translation)
\bibitem{Shuryak:2018}
          E.~Shuryak, A.V.~Turbiner,\\
          {\it Phys Rev \bf D 98} (2018) 105007 (10pp)

\bibitem{TURBINER:2005}
          A.V.~Turbiner,\\
          {\it Lett.Math.Phys. \bf 74}, 169-180 (2005)

\bibitem{TURBINER:2010}
          A.V.~Turbiner,\\
          {\it Int.Journ.Mod.Phys. \bf A25}, 647-658 (2010)

\bibitem{Baye:2015}
         D.~Baye,\\
         {\it Phys. Rep. \bf 565}, 1-108 (2015)
    
\bibitem{Tur-delValle:2021}
          A.~V.~Turbiner, J.C.~Valle,\\
          \textit{Comment on: \textit{ Uncommonly accurate energies for the general quartic oscillator}, Int. J. Quantum Chem., e26554 (2020), by P.~Okun and K.~Burke},\\
          ArXiv: 2102.09246, pp.8 (February 2021)\\
          {\it Int Journal of Quantum Chemistry \bf 122} (2021) (in press)
          
\bibitem{delValle}
          J.C.~Valle, A.~V.~Turbiner,\\
          {\it Int.Journ.Mod.Phys. \bf A34} (2019) 1950143 (43pp);
          {\it ibid \bf A35} (2020) 2050005 (44pp)


\end{thebibliography}
\end{document}